\documentclass[aps,epsfig,graphicsx,floatfix,mathbbm,a4paper]{revtex4}

\usepackage{amsmath,amsfonts,amssymb,graphics,graphicx,epsfig,color,times,bbm}

\begin{document}
\bibliographystyle{apsrev}

\newcommand{\R}{\mathbbm{R}}
\newcommand{\rr}{\mathbbm{R}}
\newcommand{\nn}{\mathbbm{N}}
\newcommand{\cc}{\mathbbm{C}}
\newcommand{\ii}{{\mathbbm{1}}}

\newcommand{\id}{{\mathbbm{1}}}
\newcommand{\flip}{ {\mathbb{F}}}

\newcommand{\tr}[1]{{\rm tr}\left[#1\right]}
\newcommand{\gr}[1]{\boldsymbol{#1}}
\newcommand{\bea}{\begin{eqnarray}}
\newcommand{\eea}{\end{eqnarray}}
\newcommand{\be}{\begin{equation}}
\newcommand{\ee}{\end{equation}}
\newcommand{\St}{{\cal S}}
\newcommand{\Ad}{{\rm Ad}}

\newcommand{\ket}[1]{|#1\rangle}
\newcommand{\bra}[1]{\langle#1|}
\newcommand{\avr}[1]{\langle#1\rangle}
\newcommand{\G}{{\cal G}}
\newcommand{\eq}[1]{Eq.~(\ref{#1})}
\newcommand{\ineq}[1]{Ineq.~(\ref{#1})}
\newcommand{\sirsection}[1]{\section{\large \sf \textbf{#1}}}
\newcommand{\sirsubsection}[1]{\subsection{\normalsize \sf \textbf{#1}}}
\newcommand{\ack}{\subsection*{\normalsize \sf \textbf{Acknowledgements}}}
\newcommand{\front}[5]{\title{\sf \textbf{\Large #1}}
\author{#2 \vspace*{.4cm}\\
\footnotesize #3}
\date{\footnotesize \sf \begin{quote}
\hspace*{.2cm}#4 \end{quote} #5} \maketitle}
\newcommand{\eg}{\emph{e.g.}~}

\newcommand{\proofend}{\hfill\fbox\\  \bigskip }

\newcommand{\proof}[1]{{\it Proof.} #1 $\proofend$}

%---------------------------------------------------------------------------

\newtheorem{theorem}{Theorem}
\newtheorem{proposition}{Proposition}

\newtheorem{lemma}{Lemma}
\newtheorem{definition}{Definition}
\newtheorem{corollary}{Corollary}
\newtheorem{example}{Example}
\newtheorem{remark}{Remark}

\title{
Classical information  capacity of a class of quantum channels}

\author{M.M.\ Wolf$^1$ and J.\ Eisert$^{2,3,4}$}

\affiliation{ 1 Max-Planck-Institut f{\"u}r Quantenoptik,
Hans-Kopfermann-Str.\ 1, 85748 Garching, Germany\\
2 Blackett Laboratory, Imperial College London, Prince Consort Rd,
London SW7 2BW, UK\\
3 Institute for Mathematical Sciences, Imperial College
London, Exhibition Rd, London SW7 2BW, UK\\
4 Institut f{\"u}r Physik, Universit{\"a}t Potsdam,
Am Neuen Palais 10, 14469 Potsdam, Germany}

\date{\today}

\begin{abstract}
We consider the additivity of the minimal output entropy and the
classical information capacity of a class of quantum channels. For
this class of channels the norm of the output is maximized
 for the output being a normalized projection. We
prove the additivity of the minimal output Renyi entropies with
entropic parameters $\alpha\in[0, 2]$, generalizing an argument by
Alicki and Fannes, and present a number of examples in detail. In
order to relate these results to the classical information
capacity, we introduce a weak form of covariance of a channel. We
then identify several instances of weakly covariant channels 
for which we can infer the additivity of the classical information capacity.
Both additivity results apply to the case of an arbitrary number
of different channels. Finally, we relate the obtained results to
instances of bi-partite quantum states for which the entanglement
cost can be calculated.
\end{abstract}

%\pacs{ }

\maketitle

\section{Introduction}

The study of capacities is at the heart of essentially any
quantitative analysis of the capabilities to  store or transmit
quantum information. This includes the case of transmission of
quantum states through noisy channels modeling decohering
transmission lines, such as fibers or waveguides in quantum
optical settings. Capacities and entropic quantities
characterizing the specifics of a given quantum channel come in
several flavors: for each resource that is allowed for, one may
define a certain asymptotic rate that can be achieved. A question
that is of key interest here -- and a notoriously difficult one --
is whether the respective quantities are generally additive. In
other words: if we encode quantum information before transmitting
it through a quantum channel, can it potentially be an advantage
to use entangled inputs over several invocations of the channel?
This question is particularly interesting for two central concepts
characterizing quantum channels: the minimal output entropy and
the classical information capacity.

The classical information capacity specifies the capability of a
noisy  channel to transmit classical information encoded in
quantum states
 \cite{OldHolevo,OldWest}.
The question of the classical information capacity is then the one
of the asymptotic efficiency of sending classical information from
sender to receiver, assuming the capability of   encoding data in
a coherent manner. This capacity is one of the central notions in
the study of quantum channels to assess their potential for
communication purposes. The minimal output entropy in turn is a
measure for the decoherence accompanied with invocations of the
channel. It specifies the minimal entropy of any output that can
 be achieved by optimizing over all channel inputs \cite{Amosov}.
 The  conjectures on general additivity of both quantities
 have been  linked to each other, in that they are either both
true or both false \cite{Shor,AB,Winter}.

It is the purpose of this paper to investigate the additivity
properties of a class of quantum channels for which the output
norm is maximized if the output state is (up to normalization) a
projection. For such channels we prove additivity of the minimal
output $\alpha$-entropies in the interval $\alpha\in[0,2]$. This
further exploits an idea going back to Alicki and Fannes in Ref.\
\cite{Fannes1} and Matsumoto and Yura in Ref.\ \cite{Matsu}. For
all weakly covariant instances of the considered channels the
additivity is shown to extend to the classical information
capacity. Both additivity results are proven for the case of an
arbitrary number of different channels.
 So on the one hand, this paper
provides several new instances of channels for which the additivity of the
minimal output entropy and the classical information capacity is
known. On the other hand, it further substantiates the conjecture
that this additivity might be generally true. Finally, following
the ideas of Ref.\ \cite{Winter}, we relate the obtained
additivity results to the additivity of the entanglement of
formation for instances of bipartite quantum states. We will begin
with an introduction of basic notions and related results in Sec.
\ref{SecPreliminaries} and the characterization of the considered
class of quantum channels in Sec. \ref{SecCharacterization}.

\section{Preliminaries}\label{SecPreliminaries}

Consider a quantum channel, i.e., a completely positive
trace-preserving map $T:\St(\cc^d)\rightarrow \St(\cc^d)$ taken to
have input and output Hilbert spaces of  dimension $d$. The
minimal output entropy of the channel, measured in terms of the
Renyi $\alpha$-entropy \cite{Renyi}, is given by
\begin{equation}\label{Min}
  \nu_{\alpha}(T):=  \inf_\rho (S_\alpha\circ T)(\rho),\quad\quad
    S_\alpha (\rho) :=   \frac{1}{1-\alpha} \log \tr{ \rho^\alpha },
\end{equation}
$\alpha\geq0$. The $\alpha$-Renyi entropies are generalizations of
the von-Neumann entropy defined as $S(\rho)=- \tr{\rho\log \rho}$,
which is  obtained in the limit $\alpha\rightarrow 1$. Therefore
we consistently define $S_1(\rho):=S(\rho)$. Physically,
$\nu_{\alpha}$ can be interpreted as a measure of decoherence
induced by the channel when acting on pure input states. The
minimal output $\alpha$-entropy is said to be additive
\cite{KRadditivityConj} if for arbitrary $N\in\nn $ 
\begin{equation}
\frac1N
\nu_{\alpha}\big(T^{\otimes N}\big) =
\nu_{\alpha}(T).\label{nuadditivity}
\end{equation}
It is known that
additivity of $\nu_{\alpha}$ does not hold in general for
$\alpha>4.79$ \cite{Werner}. For smaller values of $\alpha$,
however, no counterexample is known so far and in particular in
the interval $\alpha\in[1,2]$, where the function $x\longmapsto
x^{\alpha}$ becomes operator convex, additivity is conjectured
to hold in general.

The classical information capacity of a quantum channel, can be
inferred from its Holevo capacity \cite{OldHolevo}. The Holevo
capacity of the channel $T$ is defined as
\begin{equation}\label{Holevo}
    C(T) : = \sup \left[
    S\Big(\sum_{i=1}^n p_i T( \rho_i) \Big) - \sum_{i=1}^n  p_i (S\circ
    T) (\rho_i)\right],
\end{equation}
 $n\leq d2$,
where the supremum is taken over pure states $\rho_1,...,\rho_n\in
\St(\cc^d)$ and all probability distributions $(p_1,...,p_n)$. The
classical information capacity is according to the
Holevo-Schumacher-Westmoreland theorem \cite{OldHolevo,OldWest}
given by
\begin{equation}\label{Information}
    C_{\text{Cl}} (T) :=
    \lim_{N\rightarrow\infty} \frac{1}{N} C\big(T^{\otimes N}\big),
\end{equation}
so as the asymptotic version of the above Holevo capacity.
Unfortunately, as such, to evaluate the quantity in Eq.\
(\ref{Information}) is intractable in practice, being in general
an infinite-dimensional non-convex optimization problem. However,
in instances where one can show that
\begin{equation}\label{Cadd}
     \frac{1}{N} C\big(T^{\otimes N}\big)=C(T),
\end{equation}
for all $N\in \nn$, then Eq.\ (\ref{Holevo}) already gives the
classical information capacity. That is, to know the single-shot
quantity in  Eq.\ (\ref{Holevo}) is then sufficient to
characterize the channel with respect to its capability of
transmitting classical information. A stronger version of the
additivity statements in Eqs.\ (\ref{nuadditivity}, \ref{Cadd}) is
the one where equality is not only demanded for $N$ instances of
the same channel but for $N$ different channels
$\bigotimes_{i=1}^N T_i$. We will refer to this form of additivity
as ''strong additivity``.

The additivity of the Holevo capacity in the sense of the general
validity of Eq.\ (\ref{Cadd}) or the additivity of the minimal
output entropy is one of the key open problems in the field of
quantum information theory -- despite a significant research
effort to clarify this issue. In the case $\alpha=1$ the two
additivity statements in Eq.\ (\ref{nuadditivity}) and Eq.\
(\ref{Cadd}) were shown to be equivalent in their strong version
in the sense that if one is true for all channels (including those
with different input and output dimensions), then so is the other
\cite{Shor,AB,Winter}. For a number of channels, additivity of the
minimal output entropy for $\alpha=1$
\cite{Fannes2,Fannes1,KingUnital,KingDepolarizing,KingEntBreaking,Us,Shirokov,FHMV04}
and additivity of the Holevo capacity
\cite{FHdepolarizing,KingUnital,KingDepolarizing,ShorEntBreaking,LloydGauss,FHMV04}
are known. For integer $\alpha$, the minimal output
$\alpha$-entropy is
 more accessible than for values close to one \cite{Ruskai,GL}. Notably, for the case
 $\alpha=2$, a number of additivity statements have been derived
 \cite{p2}, and the minimal output entropy can be
 assessed with relaxation methods from global optimization \cite{Relax}.
For covariant channels, one can indeed infer the additivity of the
Holevo capacity from the additivity of the minimal output von
Neumann entropy \cite{Holevo}. In fact, as we will discuss in
Sec.\ref{SecC}, a much weaker assumption already suffices for this
implication.

A paradigmatic and well known representative of the class of
channels we consider in this paper is the Werner-Holevo channel
\cite{Werner}, which is of the form 
\begin{equation}
 T(\rho) =
\frac{\ii_d-\rho^T}{d-1}\;.
\end{equation} This channel serves as a
counter-example for the additivity of the minimal output
$\alpha$-entropy for $\alpha>4.79$. However, for $\nu_{\alpha}$
with $\alpha\in[1,2]$ and for the Holevo capacity additivity have
been proven in Refs.\ \cite{Fannes1,Matsu}. In the following we
will generalize these additivity results to a much larger class of
channels.

\section{Characterization of the
class of quantum channels}\label{SecCharacterization}

We will consider a class of channels with a remarkable property:
for this class of quantum channels one can relate the problem of
additivity of the minimal output entropy to that of another
Renyi-$\alpha$ entropy. The first key observation is the
following:

\begin{lemma}[Basic property]\label{Lemma1}
Let $T$ be a quantum channel for which
\begin{equation}
    \nu_{\alpha}(T)=\nu_{\beta}(T)
    ,\quad
    \alpha>\beta\geq0.\label{cond1}
\end{equation} Then the additivity of the
minimal output $\alpha$-entropy implies the additivity for the
minimal output $\beta$-entropy.
\end{lemma}

\proof{This statement follows immediately from the fact
$S_{\alpha}(\rho) \leq S_{\beta}(\rho)$ for all $\rho\in
\St(\cc^d)$ and all $\alpha\geq\beta\geq0$ \cite{BS}, and the
inequality chain
\begin{eqnarray}
    \nu_{\beta}(T) &= & \nu_{\alpha}(T)  =  \frac1N \nu_{\alpha}\big(T^{\otimes N}\big)
    = \frac{1}{N}
    \inf_{\rho}(S_{\alpha}\circ T^{\otimes N})(\rho) \nonumber \\
    &\leq& \frac1N  \inf_{\rho}(S_{\beta}\circ T^{\otimes N})(\rho)
    \label{ineq1},
    \end{eqnarray}
    for $N\in \nn$.
    Since on the other hand $\nu_{\beta}(T)\geq   \nu_{\beta}\big(T^{\otimes
    N}\big)/N $
equality has to hold in Eq.\ (\ref{ineq1}).}

Surprisingly, the property required in Eq.\ (\ref{cond1}) does not
restrict the channels to the extent that only trivial examples can
be found. Quite to the contrary, a fairly large class of channels
has this property. A simple example of a class of channels for
which condition (\ref{cond1}) is satisfied is the generalization
of the Werner-Holevo channel:

\begin{example}\label{FirstExample}
Consider a channel $T:\St(\cc^d)\rightarrow \St(\cc^d)$ of the
form
\begin{equation}
    T(\rho)=
    \frac{\id_d-M(\rho)}{d-1}\label{example},
\end{equation}
where $M:\St(\cc^d)\rightarrow \St(\cc^d)$ is a linear,
trace-preserving positive map (not necessarily a channel) which
has the property that there exists an input state leading to a
pure output state. Then for all $\alpha>0$
\begin{equation}
    \nu_{\alpha}(T)=\log(d-1)\label{EqLem2}.
 \end{equation}
\end{example}

\proof{Let us first note that $\rho\longmapsto
\tr{(\id_d-\rho)^\alpha}$  is convex for any $\alpha\geq 1$ and
concave for $0\leq\alpha<1$.  Hence, the sought extremum over the
convex set of all states is attained at an extreme point, i.e., a
pure state. Moreover, all pure states will give the same value.
Exploiting this together with the fact that there exists an output
under $M$ which is pure and inserting into $
    S_{\alpha}(\rho)=(\log\tr{\rho^{\alpha}})/(1-\alpha)
$ yields Eq.\ (\ref{EqLem2}).}

The class of channels in Example \ref{FirstExample} has the
property that $\nu_{\alpha}(T)$ is independent of $\alpha$ and
therefore condition (\ref{cond1}) is trivially satisfied. However,
it is not yet the most general class of channels for which
$\nu_{\alpha}$ is constant. In fact, all quantum channels
fulfilling this condition can easily be characterized. This will
be the content of the next theorem, which will make use of a Lemma
that we state subsequently. The following channels are the ones
investigated in this paper:

\begin{theorem}[Characterization of channels]\label{Thetheorem}
Let $T:\St(\cc^d)\rightarrow \St(\cc^d)$ be a quantum channel.
Then the following three statements are equivalent:
\begin{enumerate}
    \item The minimal output $\alpha$-entropy is independent of
    $\alpha$. That is, for all $\alpha>\beta\geq0$ we have
    $\nu_{\alpha}(T)=\nu_{\beta}(T)$.
    \item The channel is of the form \begin{equation}
    T(\rho)= \frac{\id_d - m M(\rho)}{d-m}\label{theform},
\end{equation}
where $M$ is a positive, linear and trace-preserving map for which
there exists an input state $\rho_0$ such that $m M(\rho_0)$ is a
projection of rank $m$.
    \item The maximal output norm $\sup_{\rho} ||T(\rho)||_{\infty}$ is attained for an output
    state being a normalized projection.
\end{enumerate}

\end{theorem}
\proof{1 $\rightarrow $ 2 : Since in general
$\rr^+\ni\alpha\longmapsto S_{\alpha}(\rho)$ is a non-increasing
function for all $\rho\in \St(\cc^d)$, there exists a state
$\rho_0$ which gives rise to the minimum in $\nu_{\alpha}$ for all
values of $\alpha$. Then, by Lemma \ref{Lemma2}, $T(\rho_0)$ has
to be a projection
 except from normalization. In particular
 $\sup_\rho ||T(\rho)||_{\infty}\leq ||T(\rho_0)||_{\infty}=1/m_0$ where
$m_0 : =\text{ rank}
 (T(\rho_0) )$. This means that the map $M_0:\St(\cc^d) \rightarrow
 \St(\cc^d)$ defined as
\begin{equation}
    M_0(\rho) : =\frac1{m_0} \id_d-T(\rho)
\end{equation}
is positive and has the property that $M_0(\rho_0)$ is except from
normalization a projection of rank $m=d-m_0$. Due to the fact that
$T$ is trace-preserving, the map $M:\St(\cc^d) \rightarrow
 \St(\cc^d)$,
\begin{equation}
    M(\rho) : =\frac{m_0}{d-m_0}M_0(\rho),
\end{equation}
is also trace-preserving. Hence, the channel $T$ has indeed a
representation of the form claimed above.\\
2 $\rightarrow $ 3 : We want to argue that $\sup_{\rho}
||\ii-mM(\rho)||_{\infty}$ is attained if $R:=m M(\rho)$ is a
projection. To this end note that $R$ is an element of the convex
set
\begin{equation}
C:=\{r\geq 0 \;|\;\tr{r}=m,\ r\leq\ii \},
\end{equation}
whose extreme points are projections of rank $m$. Remember further
that the maximum of a convex function (as the largest eigenvalue
of a positive matrix) over a closed convex set is attained at an
extreme point. When optimizing over the entire set C, the maximum
is thus attained for $R$ being a projection of rank $m$, which is
indeed accessible due to the assumed property of $M$.\\
3 $\rightarrow $ 1 : This follows immediately from
$\rr^+\ni\alpha\longmapsto S_{\alpha}$ being a non-increasing
function together with the fact that for any normalized projection
$\rho_{\text{out}}$, $S_{\alpha}(\rho_{\text{out}})=\log
{\text{rank}}\big( \rho_{\text{out}}\big)$ is independent of
$\alpha$.
 }

 \begin{lemma}\label{Lemma2}
Let $\rho\in \St (\cc^d)$ be a state for which
$S_{\alpha}(\rho)=S_{\alpha'}(\rho)$ for some
$\alpha'>\alpha\geq0$. Then $\rho$ is except from normalization a
projection and for all $\beta\geq0$
 we have
 \begin{equation}
     S_{\beta}(\rho)=\log \text{rank} (\rho).
\end{equation}
\end{lemma}
\proof{The function $\rr^+\ni\beta \longmapsto S_{\beta}(\rho)$ is
a convex and non-increasing function \cite{BS}. Hence, the
assumption in the Lemma immediately implies that
$S_{\beta}(\rho)=S_{\alpha}(\rho)=:c$ for all $\beta\geq\alpha$,
i.e.,
\begin{equation}
    \tr{\rho^{\beta}}=2^{c(1-\beta)},\quad
\end{equation}
for all $ \beta\geq\alpha$. Taking the $\beta$th root on both
sides and then the limit $\beta\rightarrow\infty$ leads to
$2^{-c}=||\rho||_{\infty}$ and thus
\begin{equation}
\tr{\big(\rho/||\rho||_{\infty}\big)^{\beta}}=||\rho||_{\infty}^{-1}.
\end{equation}
Considering again the limit $\beta\rightarrow\infty$ yields that
the multiplicity of the largest eigenvalue of $\rho$ is equal to
$||\rho||_{\infty}^{-1}$, such that $\rho$ has indeed to be a
normalized projection. }

\section{Additivity of the minimal output entropy}

For a class of channels of the form in Thm.1 we find the
additivity of the minimal output $\alpha$-Renyi entropy for
$\alpha\in[0,2] $. We exploit Lemma \ref{Lemma1} for these
channels in the simple case where $\alpha=2$ and $\beta \in[0,2
]$. What then remains to be shown is the additivity of the minimal
output $2$-entropy. This can, however, be done in the same way as
has been done in Ref.\ \cite{Fannes1} for the specific case
$M(\rho)=\rho^T$, except that more care has to be taken due to the
fact that the involved projections are not necessarily
one-dimensional.

\begin{theorem}[Strong additivity of the minimal output entropy]\label{newtheorem}
Consider channels $T_1,\ldots,T_N$ of the form in Eq.\
(\ref{theform}) such that $\bigotimes_{i=1}^N M_i$ is a positive
map. Then the minimal output $\alpha$-entropy is strongly additive
for all $\alpha\in [0,2]$, i.e.,
\begin{equation}
    \nu_{\alpha}\Big(\bigotimes_{i=1}^N T_i\Big)\ =\ \sum_{i=1}^N
    \nu_{\alpha}\big(T_i\big)\ =\ \sum_{i=1}^N \log(d_i-m_i)
\end{equation}
for $T_i:\St (\cc^{d_i})\rightarrow \St (\cc^{d_i})$ as in Eq.\
(\ref{theform}).
\end{theorem}

\proof{ We can express with $T_i (\rho) = \big(\id_{d_i} - m_i M_i
(\rho)\big)/(d_i - m_i)$ the action of the  tensor product channel
$T:=\otimes_{i=1}^N T_i$  as
\begin{equation}
    T(\rho) =
    \prod_{i=1}^N \frac{1}{d_j -m_i}
    \sum_{\Lambda\subset \{1,...,N\}}
    (\omega_\Lambda \otimes \id_{\Lambda^C})
    \prod_{k\in \Lambda}(-m_k) ,
\end{equation}
where $ \Lambda^C$ denotes the complement of $\Lambda$, $\omega:=
(M_1\otimes ...\otimes M_N) (\rho)$, and $\omega_{\Lambda}$
denotes the reduced density matrix of $\omega$ with respect to the
systems labeled with $\Lambda$. Hence, we obtain
\begin{eqnarray}
     \tr{\big(T (\rho)\big)^2 }&=&
     \prod_{i=1}^N
     \frac{1}{(d_i -m_i)^2}
     \sum_{\Lambda,\Lambda'\subset \{1,...,N\} } \prod_{k\in\Lambda}
     \prod_{l\in \Lambda'}
     (-m_k)(-m_l)
     \tr{\omega^2_{\Lambda\cap \Lambda'}}
    \prod_{k\in(\Lambda\cup\Lambda')^C} d_k
    \nonumber\\
    &=& \prod_{i=1}^N
     \frac{1}{(d_i -m_i)^2}
     \sum_{\Gamma\subset \{1,...,N\} }
     \tr{\omega^2_{\Gamma}}
     \sum_{\Delta \subset \Gamma^C}
     \sum_{\Delta' \subset \Gamma^C\backslash \Delta}
     \prod_{k\in \Delta\cup \Gamma} (-m_k)
     \prod_{l\in \Delta' \cup \Gamma} (-m_l)\nonumber\\
     &&\times
    \prod_{j \in \Gamma^C \backslash \Delta\backslash \Delta'}  d_j\nonumber\\
    &=&
     \prod_{i=1}^N
     \frac{1}{(d_i -m_i)^2}
     \sum_{\Gamma\subset \{1,...,N\} }
     \tr{\omega^2_{\Gamma} }
     \nonumber
     \prod_{k\in \Gamma } m_k^2
     \prod_{j\in \Gamma^C}
     (d_j - 2 m_j).
\end{eqnarray}
Now, exploiting the subsequently stated Lemma \ref{newlemma}, we
have $\tr{\omega_{\Gamma}^2}\leq\prod_{i\in\Gamma}m_i^{-1}$ and
thus
\begin{equation}
    \tr{\big(T (\rho)\big)^2 } \leq \prod_{i=1}^N
    \frac{1}{d_i -m_i}\;.
\end{equation}
Together with the fact that $\nu_2(T_i)=\log(d_i-m_i)$ this means
finally that we obtain 
\begin{equation}
\nu_2
(T)\geq\prod_{i=1}^N
\log(d_i-m_i)=
\sum_{i=1}^N\nu_2(T_i)
\geq\nu_2 (T),
\end{equation}
implying by Lemma \ref{Lemma1} the claimed additivity in the
entire interval $\alpha\in[0,2]$. }

\begin{lemma}\label{newlemma}
Let $M_i:{\cal S}(\mathbb{C}^{d_i})\rightarrow {\cal
S}(\mathbb{C}^{d_i})$, $i=1,\ldots,N$ be trace preserving linear
maps, for which there exist positive 
numbers $m_i\in\mathbb{N}$ such that
$\rho\mapsto \big(\ii_{d_i}\tr{\rho}-m_iM_i(\rho)\big)$ is
completely positive. If in addition $\bigotimes_{i=1}^N M_i$ is a
positive map, then
\begin{equation}\label{alw}
\forall \rho\in {\St}
({\cc}^{
{\prod_i} d_i})\;:\
    \tr{ \left(\Big(\bigotimes_{i=1}^N M_i\Big) (\rho) \right)^2}
    \leq \prod_{i=1}^N m_i^{-1}.
\end{equation}
\end{lemma}

\proof{  Let $M_i^*$ be the adjoint map defined by
$\tr{M_i^*(A)B}=\tr{AM_i(B)}$. 
Then the complete positivity
condition is equivalent to the validity of
 \begin{equation}
 \big(M_i^*\otimes\id_{d_i}\big)(P_{12}) \leq
\frac{\ii_{d_i}}{m_i}\otimes{\rm tr}_1\big[P_{12}\big]
\end{equation} 
for all positive operators 
$P_{12}\in{\cal S}(\cc^{d_i^2})$. 
In order to apply this inequality we exploit some of the properties
of the flip operator
$\mathbb{F}_d:|\Phi\rangle\otimes|\Psi\rangle\mapsto|\Psi\rangle\otimes|\Phi\rangle$
for
$|\Psi\rangle,|\Phi\rangle\in\cc^d$. 
Recall that
$\tr{A^2}=\tr{(A\otimes A)\mathbb{F}_d}$ and
$\mathbb{F}_d^{T_2}=\sum_{i,j=1}^d|i,i\rangle\langle j,j|$. Hence,
\begin{eqnarray}
 \tr{ \left(\Big(\bigotimes_{i=1}^N M_i\Big) (\rho) \right)^2}
&=&\tr{\Big[\rho\otimes\Big(\bigotimes_i
M_i\Big)(\rho)\Big]\Big[\bigotimes_i\big(M_i^*\otimes\id_{d_i}\big)
(\mathbb{F}_{d_i})\Big]}\\
&=& \tr{\Big[\rho\otimes\Big(\big(\bigotimes_i
M_i\big)(\rho)\Big)^T\Big]\Big[\bigotimes_i\big(M_i^*\otimes\id_{d_i}\big)
(\mathbb{F}_{d_i}^{T_2})\Big]}\\
&\leq& \tr{\rho\otimes\Big(\big(\bigotimes_i
M_i\big)(\rho)\Big)^T}\;\prod_j{m_j}^{-1}\\
&=& \prod_j{m_j}^{-1}\;. 
\end{eqnarray} 
}

Lemma \ref{newlemma} and therefore Thm.\ 2 require the assumption
that $\bigotimes_i M_i$ is a positive map. Although the presented
proof depends on this property, we do at present not know of any
channel of the form in Eq.\ (\ref{theform}) for which Eq.\
(\ref{alw}) is not valid. In fact, all the following examples
are such that $M_i=\Xi_i\circ\theta$, where each $\Xi_i$ is
completely positive and $\theta$ is the transposition. For all
these cases $\bigotimes_iM_i$ is evidently positive.

Obviously, Thm.\ 2 implies in particular that for any channel
$T:{\St}(\cc^d)\rightarrow {\St}(\cc^d)$ of the considered form we
have for all $\alpha\in[0,2]$ 
\begin{equation}
\frac1N
\nu_{\alpha}\big(T^{\otimes N}\big) = \nu_{\alpha}(T).
\end{equation}
 As
mentioned earlier the most prominent example of channels in the
considered class is the Werner-Holevo channel itself for which
$M(\rho)=\rho^T$. For this channel, the additivity of the minimal
output entropy has been shown in Ref.\ \cite{Matsu}, and with
inequivalent methods in Refs.\ \cite{Fannes2} and \cite{Datta}.
The following list includes further instances of
    channels for which we find
    additivity of the minimal output entropy as a
    consequence of Thm.\ \ref{newtheorem}. As stated above all
    examples are such that the corresponding $M$ is a
    concatenation of a completely positive map and the
    transposition.

\begin{example}[Stretching]\label{Stretching}
For $\omega$ being a pure state consider
    \begin{eqnarray}
        M(\rho)=  \lambda \rho^T + (1-\lambda)\omega\;,\quad
        m=1\;.
    \end{eqnarray}
    \end{example}

    Complete positivity is a consequence of this channel being
    a convex combination of the
    completely
    positive   Werner-Holevo channel and the channel $\rho\longmapsto
    (\id_d - \omega)/ (d-1)$. Obviously, $\rho_0=\omega^T$ leads to
    a normalized projection at the output.

\begin{example}[Weyl shifts]\label{Shifts} Consider the set of
unitaries $W_i= \sum_{j=1}^d |j+i  {\rm{\ mod\ } } d\rangle\langle
j|$ and take
    \begin{eqnarray}
        M(\rho) = \frac1d \sum_{i=1}^d W_i \rho^T W_i^\dagger\;,\quad
        m=1\;.
    \end{eqnarray}
\end{example}
Complete positivity of the respective channel $T$ follows from the
fact that it is a composition of the Werner-Holevo channel with
another completely positive map.  The state $\rho_0$ with $\langle
i |\rho_0 | j\rangle = 1/d$ for all $i,j=1,...,d$ is an example
for an appropriate pure input state for which $M(\rho_0)=\rho_0$.

\begin{example}[Pinching]\label{Pinching} Let $\{P_i\}$ be a set of orthogonal
projections yielding a resolution of the identity, i.e., $\sum_i
P_i=\ii_d$. Then take
    \begin{eqnarray}
        M(\rho)= \sum_{i} P_i \rho^T P_i\;,\quad m=1\;.
    \end{eqnarray}
\end{example}
Again  the respective channel $T$ is a composition of two
completely positive maps and thus itself completely positive.
Moreover, any pure state $\rho_0$ for which $\rho_0^T$ is in the
support of any $P_i$ gives rise to a normalized projection at the
output of $T$.

So far the examples were restricted to the case $m=1$. The
following examples show explicitly that all larger values of $m$
are possible as well:

\begin{example}[Casimir channel for a reducible
representation]\label{Casimir}
This example is based on a Casimir channel
$T':{\cal S}(\cc^4) \longrightarrow {\cal S}(\cc^4)$
(see Section V) for a reducible
representation of $SU(2)$,
\begin{equation}
        T'(\rho) = \sum_{i=1}^3 A_i \rho A_i^\dagger,
\end{equation}
where $A_i = (4/3)^{1/2}\pi(J_i)$, with
\begin{eqnarray}
        \pi(J_1)= \frac{i}{2}\left(
        |2\rangle\langle 3| +
        |4\rangle\langle 1| - |1\rangle\langle 4|
        - |3\rangle\langle 2|
        \right),\\
        \pi(J_2)= \frac{i}{2}\left(
        |3\rangle\langle 1| +
        |4\rangle\langle 2| - |1\rangle\langle 3|
        - |2\rangle\langle 4|
        \right),\\
        \pi(J_3)= \frac{i}{2}\left(
        |1\rangle\langle 2| +
        |4\rangle\langle 3| - |2\rangle\langle 1|
        - |3\rangle\langle 4|
        \right).
\end{eqnarray}
The operators $\pi(J_1),\pi(J_2) , \pi(J_3)$ form
generators of a four-dimensional reducible
representation of the Lie algebra of the group $SU(2)$.
As an example for $m=2$, consider the channel
\begin{equation}
        T(\rho)=  \frac{3 T' (\rho) + \rho}{4}.
\end{equation}
\end{example}
This map is clearly completely positive by
construction. We find $M$ to be
given by
\begin{equation}
        M(\rho) = \id_4/2 - T(\rho).
\end{equation}
An appropriate input $\rho_0$ for which the output is
a two-dimensional projection $M(\rho_0)=(
|3\rangle\langle 3| +|4\rangle\langle 4|
)/2$ up to
normalization is given by
\begin{equation}
        \rho_0 = \left(
        |1\rangle\langle 1 |
        + i |1\rangle\langle 4|
        - i |4\rangle\langle 1|
        +  |4\rangle\langle 4|
        \right)/2.
\end{equation}
Finally, $M$ is a positive map, as it can actually be
written as a
transposition $\theta$, followed by a completely positive map
$\Xi$, that is, $M=\Xi\circ\theta$. To show that this is
indeed the case, consider
\begin{eqnarray}
        (M\otimes { \rm{id}})(\Omega^{T_1})& =&
\frac{\id_4}{2}\otimes
        \frac{\id_4}{4}
        - \frac{3}{4} (T' \otimes { \rm{id}})(
\Omega^{T_1}) - \frac{1}{4}
        \Omega^{T_1}\geq 0,
\end{eqnarray}
where $\Omega$ is the maximally entangled state with
state vector
$|\Omega\rangle = \frac{1}{2} \sum_{i=1}^{4}
|i,i\rangle$.

\begin{example}[Shifts and pinching] Let $ W_k$ be defined as in Example
\ref{Shifts} and $K\subset\{1,\ldots,d\}$:
\begin{equation}
    M(\rho) = \frac{1}{|K|}\sum_{k\in K}\sum_{i=1}^d
    |i \rangle \langle i| \big(W_k^\dagger \rho  W_k\big)
    |i \rangle \langle i|\;,\quad m=|K|\;.
\end{equation}
\end{example}
In fact, $T$ is an entanglement-breaking channel (cf.
\cite{KingEntBreaking,ShorEntBreaking}) which can be written as
\begin{eqnarray}
    T(\rho) & =&  \frac{1}{d-|K|}
    \sum_{i=1}^d\  \langle i | \rho | i \rangle
    \sum_{k\in \{1,...,d\} \backslash K }  W_k^\dagger|i\rangle\langle i|W_k\;.
\end{eqnarray}
\begin{example}[Coarse graining]\label{Coarse}
    For $\cc^d=\cc^n\otimes\cc^D$,  consider
    \begin{equation}\label{CoarseM}
        M(\rho)= \int_{U(D)} dU \Big(\bigoplus_{i=1}^n U\Big) \rho^T \Big(\bigoplus_{i=1}^n
        U\Big)^\dagger\;,\quad m=D\;,
    \end{equation}
    where the integration is with respect to the Haar measure.
\end{example}
The averaging operation in $M$ may physically be interpreted as a
coarse graining of an operation which is only capable of resolving
$n$ blocks of size $D$ within a $d=n\cdot D$ dimensional system.
In order to prove that the above $M$ leads to an admissible and
for $n>1$ not entanglement-breaking channel, let us first note
that we may after a suitable reshuffle equivalently write \bea
M(\rho)&=&\int dU \big(\ii_n\otimes U\big)\rho^T\big(\ii_n\otimes
U\big)^\dagger \ =\  \rho^T_n\otimes\frac{\ii_D}D\;,\eea where the
tensor product is that of $\cc^d=\cc^n\otimes\cc^D$ and $\rho^T_n$
is the reduction of $\rho^T$ with respect to the first tensor
factor $\cc^n$. Obviously, $M$ is positive, trace-preserving and
for $\rho_0$ with $\langle i|\rho_0|j\rangle=1/d$ we obtain a
normalized projection of rank $D$. Complete positivity of $T$ is
equivalent to
\begin{equation}
(T\otimes\rm{id})(\Omega)\geq 0,
\end{equation}
where $ |\Omega\rangle=\frac1{\sqrt{d}}\sum_{i=1}^d|i, i\rangle$
is again 
the state vector of a maximally entangled state $\Omega$. 
Exploiting again
that the latter is related to the flip operator $\flip |i  ,
j\rangle=|j , i\rangle$ via partial transposition, i.e.,
$\Omega^{T_2}=\flip /d$, we obtain 
\begin{equation}\label{cg}
    (T\otimes{\rm{id}})(\Omega) =
    \left(\frac{\ii_{d^2}}{d}-\frac{1}{d}
    \flip_n\otimes\ii_{D^2}\right)/(d-D)\;,
\end{equation}
where $\flip_n$ is the flip operator on $\cc^n\otimes\cc^n$.
Since the latter has eigenvalues $\pm 1$, the channel defined as
above is indeed completely positive. In order to prove that $T$ is
not entanglement breaking it is sufficient to show that the
partial transpose of Eq.\ (\ref{cg}) is no longer positive, which
is true since the negative term picks up an additional factor $n$.

Finally, additivity of the minimal output entropy holds for any
channel for which there exists a pure output state, leading to a
vanishing output entropy. In this case additivity of the minimal
output entropy in the form of Eq.\ (\ref{nuadditivity}) is
evident. However, strong additivity within the considered class of
channels is still a non-trivial result.
 This applies
in particular to instances of the 3-and 4-state channels of Ref.\
\cite{34state} and the class of so-called diagonal channels, for
which strong additivity was proven recently in Ref.\
\cite{Diagonal}:

\begin{example}[Diagonal channels]\label{DiagonalChannel}
Consider $T:{\cal S}(\cc^d)\rightarrow {\cal S}(\cc^d)$ with
\begin{equation}
    T(\rho)= \sum_{k=1}^K {A_k}\rho A_k^\dagger,
\end{equation}
where $A_k$, $k=1,...,K$, are all diagonal in a distinguished
basis.
\end{example}

\section{Classical information capacity}\label{SecC}

So far we have considered the minimal output entropy of quantum
channels and their additivity properties. It turns out that for a
large subset of the considered channels, including all the
discussed Examples 3-8, one can indeed infer the additivity of the
Holevo capacity as well. On the one hand, for each covariant
instance of a quantum channel from which we know that the minimal
output entropy is additive, we can conclude that the Holevo
capacity is also additive \cite{Holevo}. For example, this
argument applies to the Werner-Holevo channel itself. One the
other hand, a quantum channel does not necessarily have to be
covariant for a very similar argument to be valid. Subsequently,
we will restate the result of Ref.\ \cite{Holevo} using
weaker assumptions. The main difference is that for a given
channel, one may exploit properties of the state for which the
output entropy is minimal. This is particularly useful in our case
at hand, where these optimal input states can always be identified
in a straightforward manner. We will first state the modified
proposition in a general way, and then apply it to the channels at
hand of the form as in Thm.\ \ref{Thetheorem}.

\begin{theorem}[Strong additivity for the classical information capacity]\label{Covariant}
Let $T:{\cal S}(\cc^d) \rightarrow {\cal S}(\cc^d)$ be a quantum
channel for which the minimal output von-Neumann entropy is
additive, and let $\{\rho_i\}$ be a set of input states for which
the minimal output entropy is achieved. If for any probability
distribution $\{p_i\}$ and $\overline{\rho}:=\sum_i p_i\rho_i$ we
have that
\begin{equation}\label{Newequation}
    (S\circ T)(\overline{\rho}) = \sup_\rho (S\circ T)(\rho)
\end{equation} holds, then the Holevo capacity $C(T)$ is additive
and the classical information capacity is given by
\begin{eqnarray}\label{second}
    C_{\rm Cl}(T) = (S\circ T)(\overline\rho)  - \nu_1(T).
\end{eqnarray} Moreover, if the assumptions are satisfied by an arbitrary number of different
channels $\{T_k\}$ among which we have strong additivity of the
minimal output entropy, then $C\big(\bigotimes_k T_k\big)=\sum_k
C(T_k)$.
\end{theorem}
\proof{Let us first consider the Holevo capacity of a single
channel. Obviously, $C(T)$ is always upper bounded by the maximal
minus the minimal output entropy. Due to the assumed properties of
the set $\{\rho_i\}$ this bound is, however, saturated and we have
\begin{eqnarray}
    C(T) &=&  \sup \left[S\biggl(\sum_{j} p_j T(\rho_j) \biggr)
    - \sum_j p_j (S\circ T)(\rho_j) \right]
    =
    (S\circ T)(\overline{\rho})  - \nu_1 (T).
\end{eqnarray}
In other words the supremum in $C(T)$ can be calculated separately
for the positive and the negative part. Now consider the
expression $C\big(\bigotimes_k T_k\big)$. If we again separate the
two suprema, then by the assumed strong additivity the maximum of
the negative part is attained for product inputs. The same is true
for the positive part, since the entropy satisfies the
sub-additivity inequality $S(\rho_{AB})\leq S(\rho_A)+S(\rho_B)$.
Hence, by evaluating the suprema separately we obtain an upper
bound which coincides with the sum of the achievable upper bounds
for the single channels.}

In practice, one is often in the position to have a channel which
is weakly covariant on an input state $\rho_0$ which minimizes the
output entropy. That is, there are unitary (not necessarily
irreducible) representations $\pi$ and $\Pi$ of a compact Lie
group or a finite group $G$ such that for all $g\in G$
\begin{equation}
T\big(\pi(g)\rho_0\pi(g)^\dagger\big)= \Pi(g)
T(\rho_0)\Pi(g)^\dagger;
\end{equation} 
in addition the image of the group
average of $\rho_0$ under $T$ is the maximally mixed state. That
is, in case of a finite group 
\begin{equation}
 \frac1{|G|}\sum_{g\in G}
\Pi(g)T(\rho_0)\Pi(g)^\dagger =\frac{\ii_d}d\;,
\end{equation} 
where we have
to replace the sum by an integral with respect to the Haar measure
if $G$ is a compact Lie group. The optimal set of states
$\{\rho_j\}$ in Thm. \ref{Covariant} is then taken to be the set
of equally distributed states $\{\pi(g)\rho_0\pi(g)^\dagger\}$
(i.e., $p_g= |G|^{-1}$ for all $g\in G$ for a finite group). In
fact, the discussed Examples 3-8 are of this weakly covariant
form.

Obviously, quantum channels which are covariant with respect to an
irreducible representation of a compact Lie group always have the
required properties. For instance for the $d$-dimensional
Werner-Holevo channel, one may take for the group $G=SU(d)$, the
defining representation $\pi$, and the conjugate representation
$\Pi$. Note, however, that the property of the channel required by
Thm.\ref{Covariant} is significantly weaker than covariance.

To construct new instances of quantum channels for which the
additivity of the classical information capacity is found, let us
consider the above mentioned examples. To start with Example
\ref{Shifts}, we know that the state $\rho_0$ with elements
$\langle i|\rho_0| j\rangle = 1/d $ for $i,j=1,...,d$ is an
optimal input. To construct an appropriate group $G$, consider the
set of unitaries,
\begin{equation}
    U_j := \sum_{l=0}^{d-1} e^{ \frac{2\pi i l j }{d} } |l\rangle\langle l|,
\end{equation}
$j=1,...,d$.
 It is straightforward to show that
\begin{eqnarray}
     T( U_j  \rho_0 U_j^\dagger)& =& U_j T(\rho_0) U_j^\dagger,
     \,\,\\
    \frac{1}{d}\sum_{j=1}^d U_j T(\rho_0) U_j^\dagger &=&
    \frac{\id_d}{d}.
\end{eqnarray}
That is, by virtue of Thm.\ \ref{Covariant} the channel in Example
\ref{Shifts} has a classical information capacity of
\begin{equation}
    C_{\rm Cl}(T)= \log(d) - \log(d-1).
\end{equation}

Example \ref{Pinching} can be treated in a similar fashion. Let us
choose the basis in which the projections are diagonal, and take
$\rho_0=|1\rangle\langle 1|$. Obviously, we have that
\begin{eqnarray}
     T( W_i \rho_0 W_i^\dagger )&=& W_i T(\rho_0) W_i^\dagger,\,\,
i=1,...,d,\\
    \frac{1}{d}\sum_{i=1}^d W_i  T(\rho_0) W_i^\dagger &=&
    \frac{\id_d}{d},
\end{eqnarray}
where the $W_i$ are again the unitary shift operators, again
forming an appropriate finite group $G$. The classical information
capacity is given by $C_{\rm Cl}(T)= \log(d) - \log(d-1)$. Note
that the same argument using shift operators, leading to a
classical information capacity of $C_{\rm Cl}(T)= \log(d)$, can be
applied to the class of diagonal channels of Example
\ref{DiagonalChannel}. This result of a maximal classical
information capacity is no surprise, however, as one can encode
classical information in a way such that information transmission
through the channel is entirely lossless.

Then, Example \ref{Casimir} is another example of a channel with
additive Holevo capacity. This becomes manifest as a consequence
of the fact that every Casimir channel \cite{Gregor} based on some
representation of $SU(2)$ is covariant under the respective
representation. Such Casimir channels are convenient building
blocks to construct a large number of channels with additive
Holevo capacity. So let us consider for $G=SU(2)$ a
$d$-dimensional representation $\pi$ of $G$ \cite{Rep}. The
generators of the associated Lie algebra are denoted with $J_k$,
$k=1,2,3$. In a mild abuse of notation, we will denote with
$\pi(J_k)$ the generators of the Lie algebra of the group $SU(2)$
in the representation $\pi$. The respective Casimir channel is
given by
\begin{equation}\label{SU2Casimir}
    T(\rho)=\frac{1}{\lambda_\pi}
    \sum_{k=1}^3 \pi(J_k) \rho \pi(J_k).
    \end{equation}
where normalization follows from the Casimir operator
\begin{equation}
       \sum_{k=1}^3 \pi(J_k)^2 = \lambda_\pi \id_d .
\end{equation}
For irreducible representations $\pi$ of $SU(2)$ we have that
$\lambda_\pi =(d-1)(d+1)/4$. The covariance of the resulting
quantum channels can be immediately deduced from the structural
constants of the Lie algebra specified as
\begin{equation}
    [ J_i, J_j] = i \varepsilon_{i,j,k} J_k, \,  \,\,\,\,
    i,j,k\in \{1,2,3\}.
\end{equation}
by making use of the exponential mapping into the group $SU(2)$.
%In particular, we have that
%\begin{eqnarray}
%    \biggl(\id_d - i \sum_{k=1}^3 x_k \pi(J_k)\biggr)
%    J_j\biggl(\id_d + i \sum_{k=1}^3 x_k \pi(J_k)\biggr)=
%    \pi(J_j) + \sum_{m=1}^3 \varepsilon_{j,l,m} x_m \pi(J_l) + O(x2)
%\end{eqnarray}
%for $x=(x_1,x_2,x_3)\in\rr3$, which can be made use of to show
%covariance.
%
%\begin{lemma}
Casimir channels $T:{\cal S}(\cc^d)\rightarrow {\cal S}(\cc^d)$
with respect to a $d$-dimensional representation $\pi$ as in Eq.\
(\ref{SU2Casimir}) are covariant in the sense that
\begin{eqnarray}
    T( \pi(g) \rho \pi(g)^\dagger) =
      \Pi(g) T(\rho) \Pi(g)^\dagger
\end{eqnarray}
for all states $\rho$, where $\Pi$ is either the defining or the
conjugate representation of $SU(2)$.
%\end{lemma}

For $d=3$, for example, we reobtain the Werner-Holevo channel.
Then, in Example \ref{Casimir} as an example of a Casimir channel
with respect to a reducible representation we find that the
channel is covariant with respect to this reducible
representation. This channel is covariant with respect to the
chosen representation $\pi$ of $SU(2)$. Moreover, we may start
from the optimal input state $\rho_0$ as specified in the example,
leading
to an output $T(\rho_0)= (|3\rangle\langle 3| +
|4\rangle\langle4| )/2$. We can generate then an
ensemble of states that averages to the maximally mixed state,
assuming the Haar measure. That is, we have that
\begin{equation}
    \int_{g\in SU(2)} dg \pi(g) T(\rho_0)
    \pi(g)^\dagger=\frac{\id_4}{4}.
\end{equation}
To be very specific, with
%\begin{equation}
    $U_x:=
    \exp( i x_2 \pi(J_2))\exp( i x_1 \pi(J_1))\exp( i x_3
    \pi(J_3))$,
%\end{equation}
$x=(x_1,x_2,x_3)\in \rr^3$, this average amounts to
\begin{equation}
    \int_0^{4\pi} dx_1
    \int_0^{\pi} dx_2
    \int_0^{2\pi} dx_3 \frac{\sin(x_2)}{16 \pi2}
    U_x T(\rho_0) U_x^\dagger=\frac{\id_4}{4}.
\end{equation}
Therefore, we again conclude that the classical information
capacity is given by $C_{\rm Cl}(T)= \log(4)- \log(2)=1$.

In a similar way, the above coarse graining channel can be shown
to exhibit an additive Holevo capacity. Here, $M(\rho)$ can be
written as in Eq.\ (\ref{CoarseM}). Therefore, the reducible
representation of $SU(n)$ corresponding to
\begin{equation}
    V\otimes \id_D,\,\,\,\,\, V\in SU(n)
\end{equation}
can be taken as the group appropriately twirling the output
resulting from the optimal input. This argument leads to an
additive Holevo capacity such that the classical information
capacity becomes
\begin{equation}
    C_{\rm Cl}(T)=
    \log(d)- \log(d-D).
\end{equation}

These examples give substance to the observation that quite many
channels of the above type can be identified for which the
classical information capacity can be evaluated.
 At this point, indeed, one may be tempted to think that {\it all}
of the above channels have an additive Holevo capacity. While we
cannot ultimately exclude this option, it is not true that Thm.\
\ref{Covariant} can be applied to all channels of the form as in
Thm.\ \ref{Thetheorem}. A simple counterexample is provided by
Example \ref{Stretching}, where only a single optimal input state
exists, namely $\rho=\omega^T$, such that Thm.\ \ref{Covariant}
cannot be applied.

%an entanglement breaking channel for which only one of the
%prepared states after measurement is pure. In this case, the image
%of all optimal input states under this channel is merely
%one-dimensional, Eq.\ (\ref{Newequation}) cannot be satisfied,
%unless all of the prepared states are pure. Another counterexample
%is in fact Example \ref{Stretching}, where Thm.\ \ref{Thetheorem}
%cannot be applied for the same reason.

\section{Note on the entanglement cost of concominant bi-partite states}

Finally, we remark on the implications of the results for the
additivity of the entanglement of formation. In Ref.\
\cite{Winter}, the additivity of weakly covariant channels has
been directly related to the additivity of the entanglement of
formation \cite{Bennett}
\begin{equation}
    E_F(\rho) = \inf \sum_{i=1}^n p_i (S \circ \text{tr}_B)(\rho_i)
\end{equation}
where the infimum is taken over all ensembles such that
$\sum_{i=1}^n p_i \rho_i=\rho$. The entanglement cost, in turn, is
the asymptotic version,
\begin{equation}
    E_C(\rho) = \lim_{N\rightarrow \infty}
    \frac{1}{N} E_F(\rho^{\otimes N}).
\end{equation}
This entanglement cost quantifies the required maximally entangled
resources to prepare an entangled state: it is the rate at which
maximally entangled states are asymptotically necessary in order
to prepare a bi-partite state using only local operations and
classical communication. In contrast to the asymptotic version of
the relative entropy of entanglement \cite{Relent}, which is known
to be different from the relative entropy of entanglement, for the
entanglement of formation no counterexample for additivity is
known. Moreover, additivity of the entanglement of formation for
all bi-partite states has been shown to be equivalent to the
strong additivity of the minimal output entropy and that of the
Holevo capacity \cite{Shor}.

For the channels considered above, the construction
 in Ref.\ \cite{Winter} can readily be applied, yielding
 further examples of states for which
 the entanglement cost is known, beyond the examples
 in Refs.\ \cite{Winter,Matsu,Cost}.
 The construction is as follows: from the
quantum channel $T:\St(\cc^d)\rightarrow \St(\cc^d)$
 one constructs a Stinespring dilation, via
 an isometry $U:\cc^d\rightarrow \cc^d\otimes \cc^{K}$ for
 appropriate $K\in \nn$.
 For any bi-partite state $\rho\in\St(\cc^d\otimes \cc^K)$
with carrier on
 %\begin{equation}
   $ {\cal K}:= U \cc^d$
 %\end{equation}
 which achieves
 \begin{equation}
    C(T)= (S\circ \text{tr}_1)( \rho ) - E_F(\rho)
 \end{equation}
 we know that
 \begin{equation}
    E_C(\rho)=E_F(\rho)= \nu_1(T).
 \end{equation}
 The following state is an example of a state with known
 entanglement cost constructed in this manner.

 \begin{example} [State with additive entanglement of formation]
 Let the state vectors from  ${\cal K}\subset \cc^4\otimes \cc^4$ be
 defined as ${\cal
 K}=\text{span}(|\psi_1\rangle,...,|\psi_4\rangle)$, with
 \begin{eqnarray}
    |\psi_1\rangle &=& \bigl(
    i( |1,4\rangle +
    |2,3\rangle-|3,2\rangle)
    +  |4,1\rangle\bigr)/2\\
     |\psi_2\rangle &=&  \bigl(
     i (
    - |1,3\rangle +
    |2,4\rangle
    +|3,1\rangle)
    + |4,2\rangle
     \bigr)/2\\
     |\psi_3\rangle &=&  \bigl( i(
    |1,2\rangle -
    |2,1\rangle+|3,4\rangle) + |4,3\rangle \bigr)/2\\
     |\psi_4\rangle &=&  
     \bigl( i(
    -|1,1\rangle -
    |2,2\rangle -|3,3\rangle) + |4,4\rangle \bigr)/2.
\end{eqnarray}
Then
%\begin{equation}
    $E_C(\rho)=E_F(\rho)= 1$,
    %\end{equation}
where
\begin{equation}
    \rho=( |\psi_1\rangle \langle\psi_1| + ...+  |\psi_4\rangle
    \langle\psi_4|)/4.
\end{equation}
 \end{example}
In just the same fashion, a large number of examples with known
entanglement cost can be constructed from the above quantum
channels.

\section{Summary and conclusions}

In this paper, we  investigated a class of quantum channels for
which the norm of the output state is maximized for an output
being a normalized projection, with respect to their additivity
properties. We introduced three equivalent characterizations of
this class of quantum channels. For all channels of this type,
which satisfy an additional (presumably weak) positivity
condition, one can infer the additivity of the minimal output von
Neumann entropy from the respective additivity in case of the
$2$-entropy. Several examples of channels of this type were
discussed in quite some detail, showing that a surprisingly large
number of quantum channels is included in the considered class.
Finally, we investigated instances of this class of quantum
channels with a weak covariance property, relating the minimal
output entropy to both the classical information capacity. This
construction gives indeed rise to a large class of channels with a
known classical information capacity.

\section{Acknowledgements}

We thank M.B.\ Ruskai and A.S.\ Holevo for valuable comments. One of
us (JE) would like to thank David Gro{\ss} for interesting
discussions. This work was supported by
 the DFG (SPP 1078),  the
European Commission (QUPRODIS IST-2001-38877), and the 
European Research Councils. 
This work benefited
from discussions during an A2 meeting, funded by the DFG (SPP
1078).

\end{document}